\documentclass[prd,aps,preprint,nofootinbib]{revtex4}
\usepackage{epsfig}
\usepackage{amsmath}

\begin{document}

\preprint{RBRC-689}

\title{Azimuthal Asymmetric Distribution of Hadrons \\Inside a Jet
at Hadron Collider}

\author{Feng Yuan}
\email{fyuan@lbl.gov} \affiliation{Nuclear Science Division,
Lawrence Berkeley National Laboratory, Berkeley, CA
94720}\affiliation{RIKEN BNL Research Center, Building 510A,
Brookhaven National Laboratory, Upton, NY 11973}

\begin{abstract}
We study the azimuthal asymmetric distribution of hadrons inside a
high energy jet in the single transverse polarized proton proton
scattering, coming from the Collins effect multiplied by the quark
transversity distribution. We argue that the Collins function in
this process is the same as that in the semi-inclusive deep
inelastic scattering. The experimental study of this process will
provide us important information on the quark transversity
distribution and test the universality of the fragmentation
functions.
\end{abstract}

\maketitle

\newcommand{\be}{\begin{equation}}
\newcommand{\ee}{\end{equation}}
\newcommand{\ben}{\[}
\newcommand{\een}{\]}
\newcommand{\beqn}{\begin{eqnarray}}
\newcommand{\eeqn}{\end{eqnarray}}
\newcommand{\Tr}{{\rm Tr} }

{\bf 1. Introduction.} Quark transversity distribution is one of
the most important quark distributions of nucleon which remains
unknown \cite{Ralston:1979ys,{Jaffe:1991kp},{Barone:2001sp}}. It
is a quark distribution when the nucleon is transversely
polarized. Unlike the polarized quark distribution in a
longitudinal polarized nucleon, the quark transversity is
difficult to measure because it is a chiral-odd distribution
\cite{Jaffe:1991kp}. For example, it can not be studied in the
inclusive deep inelastic scattering (DIS), which can only probe
the chiral-even parton distributions. The Drell-Yan lepton pair
production in $pp$ scattering can be used to study the quark
transversity distributions \cite{Ralston:1979ys,{Jaffe:1991kp}},
but have limited access to them at the collider experiment at RHIC
\cite{Martin:1997rz}.

There have been suggestions to probe the quark transversity from
other processes \cite{Barone:2001sp}. For example, in
Ref.~\cite{Collins:1992kk}, it was proposed to study the quark
transversity distributions from the semi-inclusive hadron
production in the DIS (SIDIS) process, which can couple with
another chiral-odd fragmentation function, the so-called Collins
fragmentation function, to lead to a nonzero azimuthal single spin
asymmetry (SSA). This SSA has been studied by the HERMES
collaboration at DESY \cite{Airapetian:2004tw}, and a very
interesting result on the Collins fragmentation function was found
\cite{Vogelsang:2005cs}. The Collins effect in the back-to-back
two-hadron production in $e^+e^-$ annihilation has also been
explored by the BELLE collaboration \cite{Abe:2005zx}, and a first
attempt to extract the quark transversity distribution from the
combined analysis of these two experiments has been reported
recently \cite{Anselmino:2007fs}. The interference fragmentation
function for two-hadron production has also been suggested to
study quark transversity distribution in DIS and hadronic
reactions \cite{Jaffe:1997hf}.

In this paper, we investigate the possibility to explore the quark
transversity distribution in $pp$ collision at RHIC, by studying
the azimuthal asymmetric distribution of hadrons inside a jet
\cite{Collins:1993kq}. We are interested in the hadron production
from the fragmentation of a transversely polarized quark which
inherit transverse spin from the incident nucleon through
transverse spin transfer in the hard partonic scattering processes
\cite{Stratmann:1992gu,{Collins:1993kq}}. As we show in Fig.~1, we
will study the process,
\begin{equation}
p(P_A,S_\perp)+p(P_B)\to jet (P_J)+X\to H(P_h)+X\ , \label{e1}
\end{equation}
where a transversely polarized proton with momentum $P_A$ scatters
on another proton with momentum $P_B$, and produces a jet with
momentum $P_J$ (transverse momentum $P_\perp$ and rapidity $y_1$
in the Lab frame). The three momenta of $P_A$, $P_B$ and $P_J$
will form the so-called reaction plane. Inside the produced jet,
the hadrons are distributed around the jet axes, and we are
interested in studying the azimuthal distribution of a particular
hadron $H$. This hadron will carry certain longitudinal momentum
fraction of the jet, and its transverse momentum $P_{hT}$ relative
to the jet axis will define an azimuthal angle with the reaction
plane: $\phi_h$, as shown in Fig.~1. Similarly, we can define the
azimuthal angle of the transverse polarization vector of the
incident polarized proton: $\phi_s$.

\begin{figure}[t]
\begin{center}
\includegraphics[width=9cm]{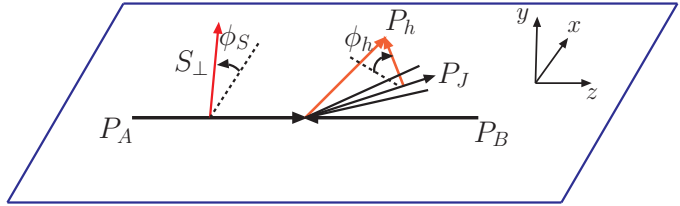}
\end{center}
\vskip -0.4cm \caption{\it Illustration of the kinematics for the
azimuthal distribution of hadrons inside a jet in $pp$
scattering.} \label{fig1}
\end{figure}

The leading order contribution to the jet production in $pp$
collision comes from $2\to 2$ sub-processes, where two jets are
produced back-to-back in the transverse plane. For the reaction
process of (1), one of the two jets shall fragment into the final
observed hadron. In this paper, we study the physics in the
kinematic region of $P_{hT}\ll P_\perp$. We assume a factorization
for this process, where we can separate the jet production from
the hadron fragmentation \cite{Collins:1981uk}. From our
calculations, we find that there exists a correlation between the
above two azimuthal angles $\phi_h$ and $\phi_s$, coming from the
quark transversity multiplied with the Collins fragmentation
function. The study of this azimuthal asymmetry will provide us
important information on the quark transversity distributions, and
will also provide a crucial test for the universality of the
Collins fragmentation function by comparing with the Collins
effects in other processes. We note that the Sivers effect
\cite{Sivers:1990fh} does not contribute to the correlation
between $\phi_h$ and $\phi_s$, because it is azimuthal symmetric
as function of $\phi_h$.

{\bf 2. Azimuthal asymmetric distribution of hadron inside a jet.}
The Collins function describes a transversely polarized quark jet
fragmenting into an unpolarized hadron, whose transverse momentum
relative to the jet axis correlates with the transverse
polarization vector of the fragmenting quark. In the fragmentation
process, the observed hadron carries certain momentum fraction
($z_h$) of the jet, and its momentum can be written as
$P_h=z_hP_J+P_{hT}$, where $P_{hT}$ is a transverse momentum
relative to the jet momentum $P_J$, i.e., $P_{hT}\cdot P_J=0$. We
notice that $P_{hT}$ may not be a transverse momentum in the Lab
frame as shown in Fig.~1. In order to observe the Collins effect
in the final state hadron distribution, the fragmenting quark has
to be transversely polarized. This can be achieved in $pp$
collision by scattering a transversely polarized quark in the
partonic process $qb\to qb$, where the final state quark $q$ can
inherit the transverse polarization from the initial state quark
$q$, and $b$ represents any other parton from the unpolarized
proton. Thus, this Collins effect will depend on the quark
transversity distribution of the transversely polarized proton in
the initial state. The contribution to the transverse-spin
dependent cross section for the process (1) from the $2\to 2$
subprocess $qb\to qb$ can be calculated, and we find that
\begin{eqnarray}
\frac{d\sigma(S_\perp)}{d{\cal P.S.}}=\sum_{b=q,g}x'f_b(x')
x\delta q_T(x)\delta \hat q(z_h,P_{hT})\frac{\epsilon^{\alpha\beta}S_\perp^\alpha}{M_h}\nonumber\\
\times \left[P_{hT}^\beta-\frac{P_B\cdot P_{hT}}{P_B\cdot
P_J}P_{J}^\beta\right]\times H_{qb\to qb}^{\rm Collins}\ ,
\end{eqnarray}
where $d{\cal P.S.}={dy_1dy_2dP_\perp^2dzd^2P_{hT}}$ represents
the phase space for this process, $y_1$ and $y_2$ are rapidities
for the jet $P_J$ and the balancing jet, respectively. $P_\perp$
is the jet's transverse momentum. The final observed hadron's
kinematic variables $z_h$ and $P_{hT}$ are defined above. Here,
$x$ and $x'$ are the momentum fractions carried by the quark
``$q$" and parton ``$b$" from the incident polarized and
unpolarized nucleons. $f_b$ is the parton distribution for ``$b$",
$\delta q_T(x)$ (also noted as $\delta q$, $h_{1q}$ and $\Delta_T
q$ in the literature) is the quark transversity distribution, and
$\delta \hat q$ the Collins fragmentation function
\cite{Collins:1992kk} (also noted as $\Delta \hat D$ or
$H_1^\perp$ in the literature). The quark transversity and the
Collins function follow the convention used in
\cite{Ji:2004wu,{Vogelsang:2005cs}}, which is different from the
so-called Trento convention for the Collins function: $\hat \delta
q\equiv -H_{1}^\perp/z_h$. $H_{qb\to qb}^{\rm Collins}$ is the
hard factor for the partonic channel $qb\to qb$. Because the quark
chirality is conserved, we only have the following channels
contributing to the above cross section: $qq'(\bar q')\to qq'(\bar
q')$, $qq\to qq$ and $qg\to qg$, and those with the anti-quark
transversity. The hard factors are
\begin{eqnarray}
H_{qq'\to qq'}^{\rm Collins}&=&H_{q\bar q'\to q\bar q'}^{\rm
Collins}=\frac{\alpha_s^2\pi}{\hat
s^2}\frac{N_c^2-1}{4N_c^2}\frac{4\hat s\hat u}{-\hat t^2} \ ,
\nonumber\\
H_{qq\to qq}^{\rm Collins}&=&\frac{\alpha_s^2\pi}{\hat
s^2}\frac{N_c^2-1}{4N_c^2}\left[\frac{4\hat s\hat u}{-\hat
t^2}-\frac{1}{N_c}\frac{4\hat s}{-\hat t}\right] \ ,
\nonumber\\
H_{qg\to qg}^{\rm Collins}&=&\frac{\alpha_s^2\pi}{\hat
s^2}\left[\frac{N_c^2-1}{N_c^2}+\frac{1}{2}\frac{4\hat s\hat
u}{-\hat t^2}\right] \ ,
\end{eqnarray}
where $\hat s$, $\hat t$, and $\hat u$ are the usual partonic
Mandelstam variables. The hard factors for the partonic channels
associated with the antiquark transversity are the same as the
above. Not surprisingly, these hard factors are exactly the same
as those calculated for the transverse spin transfer in the hard
partonic processes \cite{Stratmann:1992gu,{Collins:1993kq}}.

With the kinematics shown in Fig.~1, the above differential cross
section can be further simplified, because $
\epsilon^{\alpha\beta}S_\perp^\alpha\left[P_{hT}^\beta-\frac{P_B\cdot
P_{hT}}{P_B\cdot
P_J}P_{J}^\beta\right]=|S_\perp||P_{hT}|\sin(\phi_h-\phi_S)$,
where $\phi_S$ and $\phi_h$ are the azimuthal angles defined
above. The differential cross sections for the hadron distribution
in a jet can then be summarized as
\begin{equation}
\frac{d\sigma}{d{\cal P.S.}}= \frac{d\sigma_{UU}}{d{\cal P.S.}}+
|S_\perp|\frac{|P_{hT}|}{M_h}\sin(\phi_h-\phi_s)
\frac{d\sigma_{TU}}{d{\cal P.S.}}\ ,\label{e8}
\end{equation}
where $d\sigma_{UU}$ and $d\sigma_{TU}$ are the the spin-averaged
and single-transverse-spin dependent cross section terms,
respectively. They are defined as
\begin{eqnarray}
\frac{d\sigma_{UU}}{d{\cal P.S.}}&=&\sum_{a,b,c}x'f_{b}(x')xf_a(x)
D_c^h(z,P_{hT})H_{ab\to
cd}^{uu} \ ,\nonumber\\
\frac{d\sigma_{TU}}{d{\cal P.S.}}&=&\sum_{b,q}x'f_{b}(x')x\delta
q_T(x) \delta \hat q(z,P_{hT})H_{qb\to qb}^{\rm Collins} \ ,
\end{eqnarray}
where the hard factors for the spin-averaged cross sections are
identical to the differential partonic cross sections: $H_{ab\to
cd}^{uu}=d\hat\sigma_{ab\to cd}^{uu}/d\hat t$.

{\bf 3. Universality of the Collins fragmentation function.} The
Collins asymmetry has been studied by the HERMES collaboration in
the SIDIS process \cite{Airapetian:2004tw}, and by the BELLE
collaboration in $e^+e^-$ annihilation \cite{Abe:2005zx}. In order
to use their constraints to predict the asymmetry of
Eq.~(\ref{e8}) in the $pp$ collisions, we will assume the
universality of the Collins functions in these processes. In
Ref.~\cite{Metz:2002iz}, it has been shown that the Collins
function is universal between the DIS and $e^+e^-$ annihilation
processes, which was later argued on a more general ground from
the factorization property of the relevant processes
\cite{{Collins:2004nx}}.

\begin{figure}[t]
\begin{center}
\includegraphics[width=10cm]{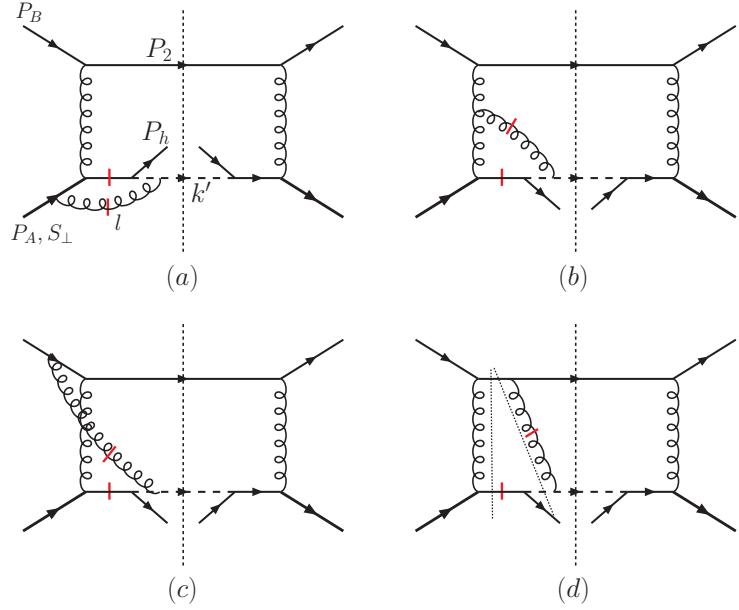}
\end{center}
\vskip -0.4cm \caption{\it Gluon exchange diagrams contributions
to the Collins asymmetry in $pp$ collisions. The short bars
indicate the pole contributions to the phase needed for a
non-vanishing SSA. The additional two cuts in (d) cancel out each
other.} \label{fig2}
\end{figure}

In the following, we will extend the universality discussion to
our case, and argue that the Collins function of the hadron
production in a jet fragmentation in $pp$ collision will be the
same as that in the SIDIS process. The way we demonstrate the
universality is similar to the model calculation in
\cite{Metz:2002iz}. The conclusion, however, does not depend on
the model. As we show in Fig.~2, we use a generic model to couple
the find state hadron to the fragmenting quark
\cite{Collins:1992kk}, which is produced, for example, from the
partonic process $q(S_\perp)q'\to q(s_\perp)q'$, with both the
initial and final state quarks transversely polarized. We focus on
the discussions for this particular channel, and all other
channels follow accordingly \cite{future}.

For the single-transverse-spin dependent cross section from the
Collins effect, we need to generate a phase from the scattering
amplitudes to have a non-vanishing SSA. If the phase comes from
the vertex associated with the fragmenting quark and the final
state hadron \cite{Collins:1992kk}, or from the dressed quark
propagator \cite{Amrath:2005gv}, it is easy to argue the
universality of the Collins function between our process and the
SIDIS/$e^+e^-$ process, because they are the same. The main issue
of the universality discussion concerns the extra gluon exchange
contribution between the spectator and hard partonic part
\cite{Metz:2002iz}. For example, in our case, because the hadron
is colorless while the quark is colored, the remanet in the
fragmentation process will be also colored. Thus the gluon
exchanges between the remanet and the other parts of the
scattering amplitudes become essential. In Fig.~2, we have shown
all these interactions, including the gluon attachments to the
incident quarks (a,c), and final state balancing quark (d) and the
internal gluon propagator (b). Although these diagrams are much
more complicated than that discussed in \cite{Metz:2002iz} for
SIDIS and $e^+e^-$ processes, we can still study their
contributions, by classifying different momentum regions for the
exchanged gluons. The dominant contribution to the fragmentation
function comes from the kinematic region where the exchanged gluon
is parallel to the final state hadron \cite{Collins:1989gx}.
Otherwise, their contributions will be power suppressed in the
limit of $P_{hT}\ll P_\perp$
\cite{{Collins:1989gx},future,Qiu:2007ar}. For these collinear
gluon interactions, we can use Ward identity and eikonal
approximation to sum them together to form the gauge link in the
definition of the fragmentation function
\cite{Collins:1989gx,future}.

The contributing phases of the diagrams in Fig.~2 come from the
cuts through the internal propagators in the partonic scattering
amplitudes \cite{{Brodsky:2002cx},{Metz:2002iz}}. In Fig.~2, we
labelled these cut-poles by short bars in the diagrams. From our
calculations, we find that all these poles come from a cut through
the exchanged gluon and the fragmenting quark in each diagram, and
all other contributions either vanish or cancel out each other
\cite{future}. For example, in Fig.~2(d), we show two additional
cuts, which contribute however opposite to each other and cancel
out completely. Therefore, by using the Ward identity at this
particular order, the final results for all these diagrams will
still sum up together into a factorized form as shown in Fig.~3,
where the cross section is written as the hard partonic cross
section for $q(S_\perp)q'\to q(s_\perp)q'$ subprocess multiplied
by a Collins fragmentation function \cite{future}. The exchanged
gluon in Fig.~2 is now attaching to a gauge link from the
fragmentation function definition \cite{Collins:1981uk}. The
Collins fragmentation function can be calculated from this
diagram, and it will not depend on the gauge link direction
\cite{Amrath:2005gv}. Clearly, this demonstrates the universality
property \cite{Metz:2002iz,{Collins:2004nx},future}, and the
Collins function for our process will be the same as that in the
SIDIS and $e^+e^-$ annihilation processes. We emphasize this
conclusion is model-independent. This observation is very
different from that for the parton distributions, where the
initial/final state interactions from the gluon exchange between
the spectator and the active quark change the normal universality
property for the so-called naive time-reversal-odd parton
distributions, for which we will have opposite signs for the SIDIS
and Drell-Yan processes
\cite{Brodsky:2002cx,{Collins:2002kn},{Ji:2002aa}}.

\begin{figure}[t]
\begin{center}
\includegraphics[width=10cm]{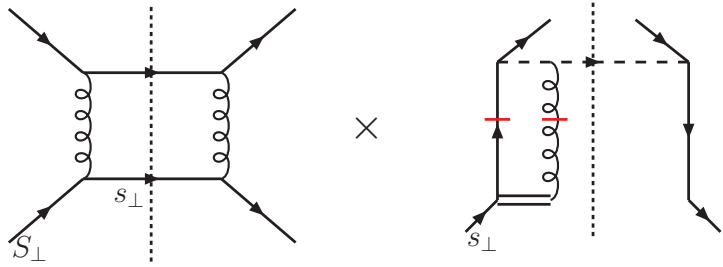}
\end{center}
\vskip -0.4cm \caption{\it Factorize the contributions from
Fig.~\ref{fig2} into the hard partonic cross section multiplied by
the universal Collins fragmentation function. The short bars
indicate the pole contribution to the Collins function.}
\label{fig3}
\end{figure}

With the universality property of the Collins fragmentation
function, we can predict the azimuthal asymmetry of the hadron
production inside a jet in the single transversely polarized $pp$
collisions from the knowledge of the Collins function in other
processes \cite{{Airapetian:2004tw},Abe:2005zx}. One interesting
asymmetry is calculated from Eq.~(\ref{e8}), by integrating over
$|P_{hT}|$ while keeping the azimuthal dependence
$\sin(\phi_h-\phi_s)$, and the cross section can be written as
$d\sigma=d\sigma_{UU}\left(1+A_N\sin(\phi_h-\phi_s)\right)$. The
asymmetry $A_N$ is defined as
\begin{equation}
A_N=\frac{\int dy_2 \sum_{qb}x'f_b(x')x\delta q_T(x) \delta \hat
q^{(1/2)}(z_h)H_{qb\to qb}^{\rm Collins}} {\int dy_2
\sum_{abc}x'f_b(x')xf_a(x) D_c^h(z_h)H_{ab\to cd}^{\rm UU}} \
,\label{e10}
\end{equation}
where $\delta\hat q^{(1/2)}(z_h)$ is the so-called $1/2$-moment of
the Collins function,
\begin{equation}
\delta\hat q^{(1/2)}(z_h)=\int
d^2P_{hT}\frac{|P_{hT}|}{M_h}\delta\hat q(z_h,P_{hT}) \ .
\end{equation}
The above functions for the pions have been fit to the HERMES data
\cite{Airapetian:2004tw} in terms of unpolarized fragmentation
functions: $\delta \hat u^{\pi^+(1/2)}=\delta \hat
d^{\pi^-(1/2)}=C_fz(1-z) D_u^{\pi^+}$; $\delta \hat
u^{\pi^-(1/2)}=\delta \hat d^{\pi^+(1/2)}=C_u z(1-z)D_u^{\pi^-}$,
where $D_u^{\pi^+}$ and $D_u^{\pi^-}$ are also called favored and
un-favored fragmentation functions for pions, respectively, and
the coefficients are found as $C_f=-0.29$ and $C_u=0.56$
\cite{Vogelsang:2005cs}, by using the quark transversity
distributions parameterized in \cite{Martin:1997rz}. We noticed
that the Collins asymmetries found in SIDIS and $e^+e^-$
annihilation are also consistent with each other
\cite{Anselmino:2007fs}. In Fig.~4, we plot the asymmetries of
Eq.~(\ref{e10}) for the charged and neutral pions at RHIC as
functions of rapidity $y_1$ and transverse momentum $P_\perp$ of
the jet. From these plots, we find that the asymmetries for the
charged pions are sizable at forward rapidity region, whereas that
for the neutral pion is very small due to the strong cancellation
between the favored and unfavored Collins functions in the fit to
the HERMES data \cite{Vogelsang:2005cs}. It will be interested to
compare with the predictions based on the quark transversity and
Collins fragmentation functions obtained in
\cite{Anselmino:2007fs}.

\begin{figure}[t]
\begin{center}
\includegraphics[height=5.0cm,angle=0]{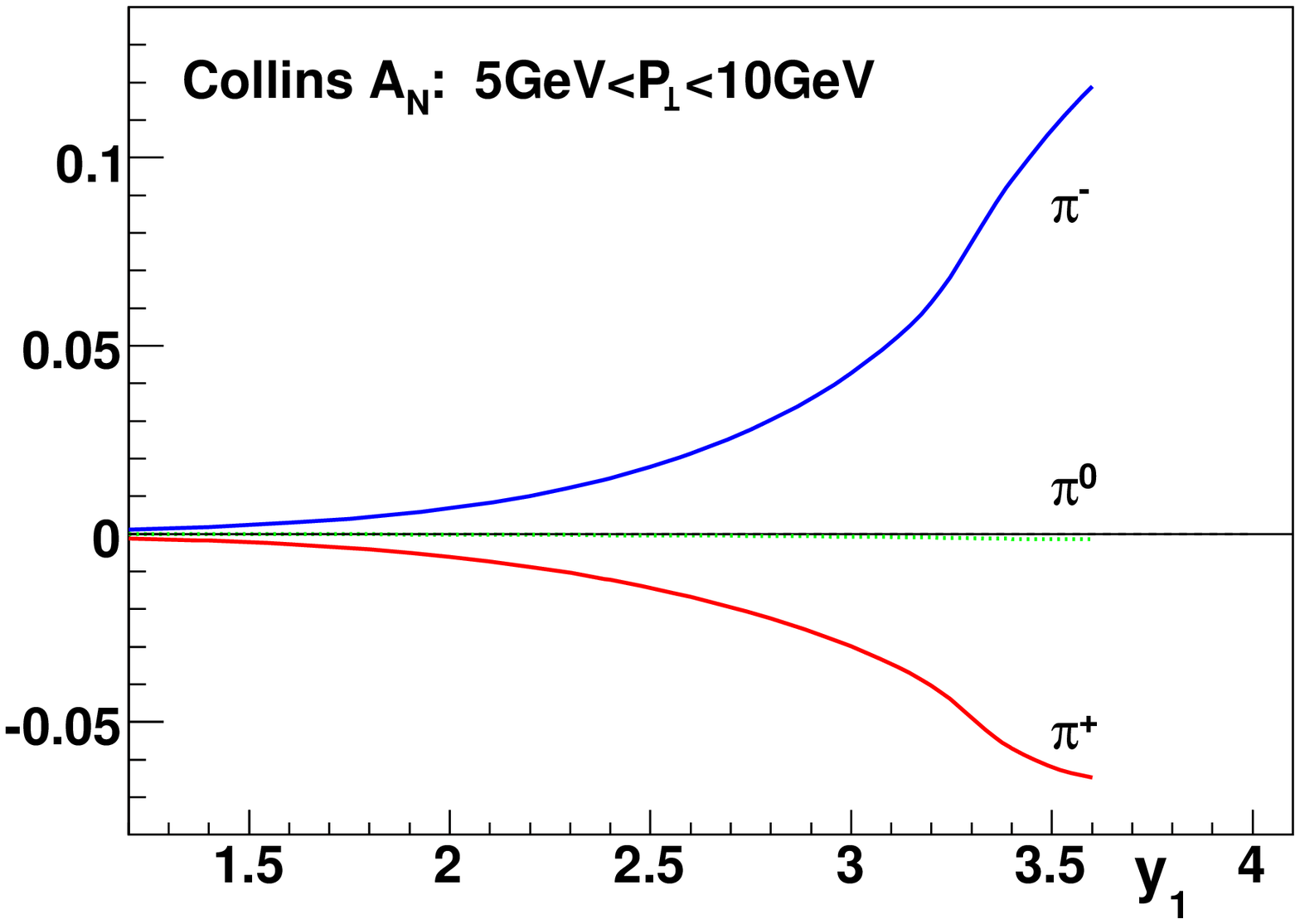}
\includegraphics[height=5.0cm,angle=0]{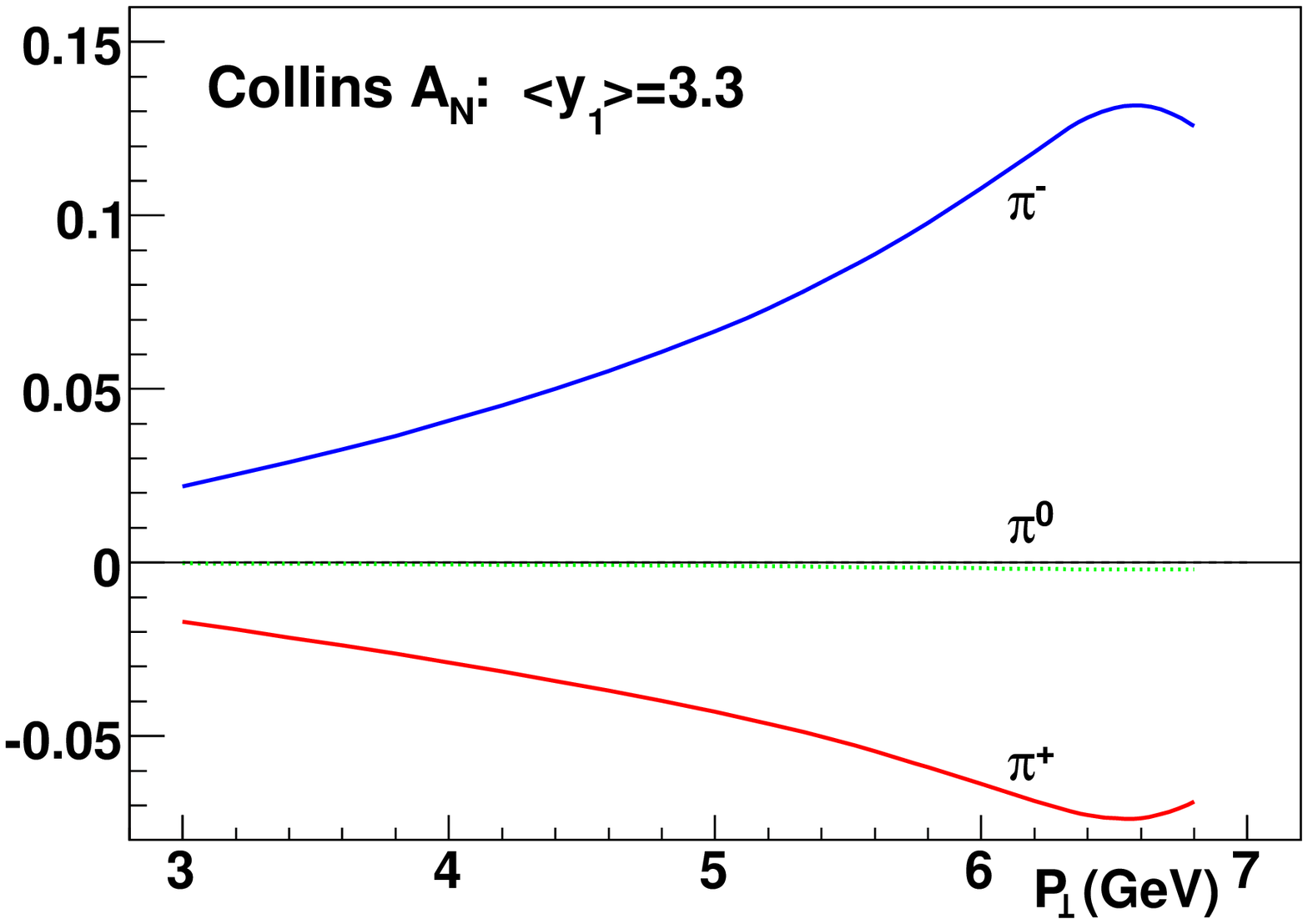}
\vspace*{-0.2cm} \epsfysize=4.2in
\end{center}
\caption{\it Collins SSAs calculated from Eq.~(\ref{e8}) for pions
in a jet in $pp$ collision at $\sqrt{s}=200GeV$ at RHIC: left
panel as functions of the jet rapidity $y_1$; right panel as
functions of the jet transverse momentum $P_\perp$.}
\end{figure}

{\bf 4. Summary.} In this paper, we have proposed to study the
quark transversity distribution by measuring the azimuthal
asymmetry of hadron production inside a jet in the
single-transverse-spin dependent $pp$ collisions at RHIC. We have
argued the universality of the Collins function between this and
other processes. By using the information on the Collins functions
from HERMES experiment, we predicted the azimuthal asymmetries for
charged and neutral pions in a jet at RHIC, and the SSAs for the
charged pions are found sizable in forward region of the polarized
proton beam. The experimental study of these asymmetries will be
crucial to test the universality of the Collins fragmentation
function, and provide us important information on the quark
transversity distributions.

A number of extensions can be followed based on our study. One of
the important issues is the QCD factorization. In our calculation,
we assumed the factorization works, and demonstrated the
universality of the Collins function in a model calculation. It
will be crucial to show this property in a real QCD framework.
Another important aspect associated with the Collins function is
the connection with the quark-gluon correlation contribution in
the fragmentation process \cite{Kanazawa:2000hz,{Boer:2003cm}}. We
reserve these further studies in a future publication, together
with a detailed derivation of our results in this paper.

We thank L.~Bland, G.~Bunce, M.~Chiu, J.~Collins, L.~Gamberg,
R.~Jaffe, X.~Ji, A.~Metz, J.~Qiu, and W.~Vogelsang for useful
discussions. This work was supported in part by the U.S.
Department of Energy under grant contract DE-AC02-05CH11231. We
are grateful to RIKEN, Brookhaven National Laboratory and the U.S.
Department of Energy (contract number DE-AC02-98CH10886) for
providing the facilities essential for the completion of this
work.

\end{document}